\documentclass[twoside,10pt,a4paper]{newFNLstyle}
\usepackage{graphics}
\usepackage{cite}
\usepackage{epsfig}

\def\w0{\omega_0}

\begin{document}

\volnumpagesyear{0}{0}{000--000}{2005}
\dates{\today}{}{}

\title{NONMONOTONIC PATTERN FORMATION IN THREE SPECIES LOTKA-VOLTERRA
SYSTEM WITH COLORED NOISE}

\authorstwo{A. Fiasconaro$^*$, D. Valenti and B. Spagnolo}
\affiliationtwo{Dipartimento di Fisica e Tecnologie Relative and
INFM, Group of Interdisciplinary Physics\footnote {Electronic
address: http://gip.dft.unipa.it},} \mailingtwo{Universit\`a di
Palermo, Viale delle Scienze pad. 18, I-90128 Palermo, Italy \\
$^*$afiasconaro@gip.dft.unipa.it}


\maketitle

\markboth{A. Fiasconaro, D. Valenti and B. Spagnolo}{Pattern
formation in three species Lotka-Volterra system with colored
noise}

\pagestyle{myheadings}


\begin{abstract}
A coupled map lattice of generalized Lotka-Volterra equations in
the presence of colored multiplicative noise is used to analyze
the spatiotemporal evolution of three interacting species: one
predator and two preys symmetrically competing each other. The
correlation of the species concentration over the grid as a
function of time and of the noise intensity is investigated. The
presence of noise induces pattern formation, whose dimensions show
a nonmonotonic behavior as a function of the noise intensity. The
colored noise induces a greater dimension of the patterns with
respect to the white noise case and a shift of the maximum of its
area towards higher values of the noise intensity.
\end{abstract}

\section{Introduction}
The addition of noise in mathematical models of population
dynamics can be useful to describe the observed phenomenology in a
realistic and relatively simple form. This noise contribution can
give rise to non trivial effects, modifying sometimes in an
unexpected way the deterministic dynamics. Examples of noise
induced phenomena are stochastic resonance, noise delayed
extinction, temporal oscillations and noise-induced pattern
formation \cite{Val,Lutz,Sancho,Garcia-Sancho,Katja}. Biological
complex systems can be modelled as open systems in which
interactions between the components are nonlinear and a noisy
interaction with the environment is present \cite{Ciuchi}.
Recently it has been found that nonlinear interaction and the
presence of multiplicative noise can give rise to pattern
formation in population dynamics of spatially extended systems
\cite{Spa,Spa1,Ale1}. The real noise sources are correlated and
their effects on spatially extended systems have been investigated
in Refs. \cite{Sancho} (see cited Refs. there) and
\cite{Garcia-Sancho}. In this paper we study the spatio-temporal
evolution of an ecosystem of three interacting species: two
competing preys and one predator, in the presence of a colored
multiplicative noise. We find a nonmonotonic behavior of the
average size of the patterns as a function of the noise intensity.
The effects induced by the colored noise, in comparison with the
white noise case \cite{Ale1}, are: (i) pattern formation with a
greater dimension of the average area, (ii) a shift of the maximum
of the area of the patterns towards higher values of the
multiplicative noise intensity.
\section{The model}
To describe the dynamics of our spatially distributed system, we
use a coupled map lattice (CML) \cite{Ale1,cml} with a
multiplicative noise

\begin{eqnarray}
x_{i,j}^{n+1} & = & \mu x_{i,j}^n (1 - x_{i,j}^n-\beta^n
y_{i,j}^n-\alpha z_{i,j}^n)+ x_{i,j}^n X_{i,j}^n + D\sum_p
(x_{p}^n-x_{i,j}^n),
\nonumber \\
y_{i,j}^{n+1} & = & \mu y_{i,j}^n (1 - y_{i,j}^n-\beta^n
x_{i,j}^n-\alpha z_{i,j}^n)+ y_{i,j}^n Y_{i,j}^n +D\sum_p
(y_{p}^n-y_{i,j}^n),
\nonumber \\
z_{i,j}^{n+1} & = & \mu_z z_{i,j}^n
[-1+\gamma(x_{i,j}^n+y_{i,j}^n)] + z_{i,j}^n Z_{i,j}^n + D\sum_p
(z_{p}^n-z_{i,j}^n),
 \label{eqset}
\end{eqnarray}
where $x_{i,j}^n$, $y_{i,j}^n$ and $z_{i,j}^n$ are respectively
the densities of preys $x$, $y$ and the predator $z$ in the site
$(i,j)$ at the time step $n$. Here $\alpha$ and $\gamma$ are the
interaction parameters between preys and predator, $D$ is the
diffusion coefficient, $\mu$ and $\mu_z$ are scale factors.
$\sum_{p}$ indicates the sum over the four nearest neighbors in
the map lattice. $X(t), Y(t), Z(t)$ are Ornstein-Uhlenbeck
processes with the statistical properties

 \begin{equation}
  \langle \chi(t) \rangle =  0, \:\:\:\:\: \langle \langle \chi(t)
  \chi(t+\tau) \rangle = \frac{q}{2 \tau_c}
   e^{-\tau / \tau_c},  \label{meanou}
 \end{equation}
and
 \begin{equation}
   \langle X^n_{i,j} Y^m_{i,j} \rangle  = \langle X^n_{i,j} Z^m_{i,j}\rangle  =
   \langle Y^n_{i,j} Z^m_{i,j} \rangle = 0 \;\;\; \forall \; n,m,i,j
 \end{equation}
where $\tau_c$ is the correlation time of the process, $q$ is the
noise intensity, and $\chi(t)$ represents the three continuous
stochastic variables ($X(t), Y(t), Z(t)$), taken at time step $n$.
The boundary conditions are such that no interaction is present
out of lattice. Because of the environment temperature, the
interaction parameter $\beta(t)$ between the two preys can be
modelled as a periodical function of time

\begin{equation}
 \beta(t)=1 + \epsilon + \eta cos(\omega t).
 \label{betat}
\end{equation}
 Here $\eta = 0.2$, $\omega =
\pi 10^{-3}$ and $\epsilon=-0.1$. The interaction parameter
$\beta(t)$ oscillates around the critical value $\beta_c=1$ in
such a way that the dynamical regime of Lotka-Volterra model for
two competing species changes from coexistence of the two preys
($\beta<1$) to exclusion of one of them ($\beta>1$). The
parameters used in our simulations are the same of \cite{Ale1}, in
order to compare the results with the white noise case.
Specifically they are: $\mu = 2$; $\alpha = 0.03$; $\mu_z = 0.02$,
$\gamma = 205$ and $D = 0.1$. The noise intensity $q$ varies
between $10^{-11}$ and $10^{-2}$. With this choice of parameters
the intraspecies competition among the two prey populations is
stronger compared to the interspecies interaction preys-predator
($\beta \gg \alpha$), and both prey populations can therefore
stably coexist in the presence of the predator \cite{Baz}. To
evaluate the species correlation over the grid we consider the
correlation coefficient $r^n$ between a couple of them at the step
$n$ as

\begin{equation}
r^n = \frac{\sum_{i,j}^N (w_{i,j}^n - \bar{w}^n) (k_{i,j}^n -
\bar{k}^n)}{\left[\sum_{i,j}^N (w_{i,j}^n - \bar{w}^n)^2
\sum_{i,j}^N (k_{i,j}^n - \bar{k}^n)^2 \right]^{1/2}},
\label{r}
\end{equation}
where $N$ is the number of sites in the grid ($100\mathrm{x}100$),
the symbols $w^n, k^n$ represent one of the three species
concentration $x, y, z$, and $\bar{w}^n,\bar{k}^n$ are the mean
values of the same quantities in all the lattice at the time step
$n$. From the definition (\ref{r}) it follows that $ -1 \leq r^n
\leq 1$.

\section{Colored Noise effects}
We quantify our analysis by considering the maximum patterns,
defined as the ensemble of adjoining sites in the lattice for
which the density of the species belongs to the interval $[3/4 \;
max, max]$, where $max$ is the absolute maximum of density in the
specific grid. The various quantities, such as pattern area and
correlation parameter, have been averaged over 50 realizations,
obtaining the mean values below reported. We evaluated for each
spatial distribution, in a temporal step and for a given noise
intensity value, the following quantities referring to the maximum
pattern (MP): mean area of the various MPs found in the lattice
and correlation $r$ between two preys, and between preys and
predator.

\begin{figure}[htbp]
 \begin{center}
  \vskip -0.5cm
   \includegraphics[height=8cm]{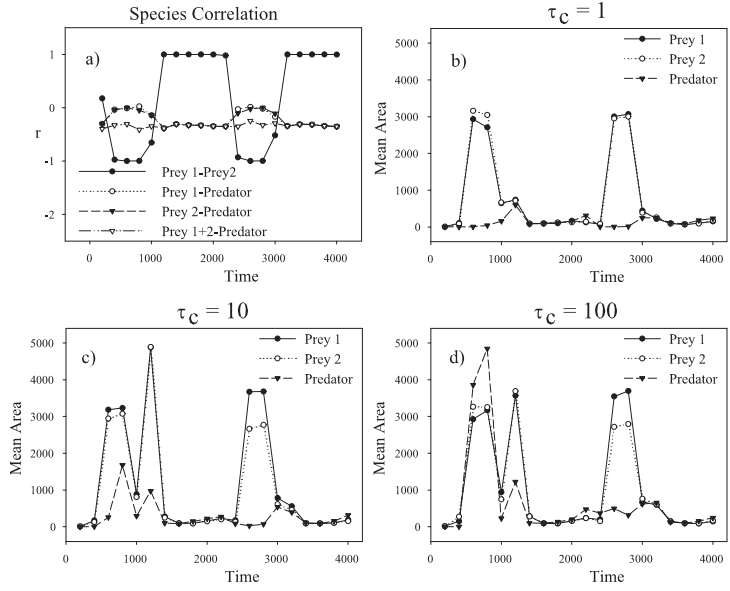}
  \vskip -0.3cm
\caption{(a) The correlation coefficient between preys and between
preys and predator as a function of time; (b-d) Mean area of the
maximum patterns of the species as a function of time, for three
values of correlation time $\tau_c = 1, 10, 100$  and for $q =
10^{-4}$. The correlation plot (a) is quite the same for all the
$\tau_c$ investigated.}
  \vskip -0.5cm
  \label{cor}
 \end{center}%
\end{figure}

From the deterministic analysis we observe: (i) for $\epsilon < 0$
($\beta < 1$) a coexistence regime of the two preys, characterized
in the lattice by a strong correlation between them and the
predator lightly anti-correlated with the two preys; (ii) for
$\epsilon > 0$ ($\beta > 1$) wide exclusion zones in the lattice,
characterized by a strong anti-correlation between preys. Because
of the periodic variation of the interaction parameter $\beta(t)$,
an interesting activation phenomenon for $\epsilon < 0$ takes
place: the two preys, after an initial transient, remain strongly
correlated for all the time, in spite of the fact that the
parameter $\beta(t)$ takes values greater than $1$ during the
periodical evolution. We focus on this dynamical regime to analyze
the effect of the noise. We found that the noise acts as a trigger
of the oscillating behavior of the species correlation $r$ giving
rise to periodical alternation of coexistence and exclusion
regime. Even a very small amount of noise is able to destroy the
coexistence regime periodically in time. This gives rise to a
periodical time behavior of the correlation parameter $r$, with
the same periodicity of the interaction parameter $\beta(t)$ (see
Eq.(\ref{betat})), which turns out almost independent of the noise
intensity and of the correlation time $\tau_c$ (see
Fig.\ref{cor}(a)). This periodicity reflects the periodical time
behavior of the mean area of the patterns. A nonmonotonic behavior
of the pattern area as a function of time is observed for all
values of noise intensity investigated. This behavior becomes
periodically in time for lower values of noise intensity, when
higher values of correlation time $\tau_c$ are considered. In
Figs.\ref{cor}(b-d) we show the time evolution of the mean area of
the maximum patterns, for $q = 10^{-4}$ and for three values of
correlation time, namely $\tau_c = 1, 10, 100$. The periodicity of
the nonmonotonic behavior of the area of MPs is clearly observed.

To analyze the noise induced pattern formation we focus on the
correlation regime between preys $r_{12} = 1$, where pattern
formation appears. In fact when the preys are highly
anticorrelated with species correlation parameter $r_{12} = -1$, a
big clusterization of preys is observed, with large patches of
preys enlarging to all the available space of the lattice. This
scenario, observed also in the white noise case \cite{Ale1}, is
confirmed by the analysis of the time series of the species. These
large patches appear, in the anticorrelation regime corresponding
to the exclusion regime of the two preys, with smooth contours and
low intensity of species density for lower noise intensities and
higher correlation time values.

\begin{figure}[htbp]
 \begin{center}
  \includegraphics[height=6cm]{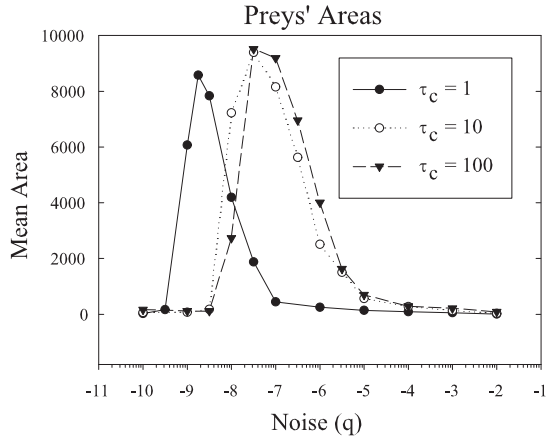}
 \vskip -0.3cm
\caption{Semi-Log plot of the mean area of the maximum patterns
for all species as a function of noise intensity, at iteration
step $1400$ for the three correlation time here reported. See the
text for the values of the other parameters.}
 \vskip -0.5cm
  \label{aree}
 \end{center}
\end{figure}

 The study of the area of the pattern formation as a function of
 noise intensity with colored noise shows two main effects: 1) the
increase of the pattern dimension and 2) a shift of the maximum
toward higher values of the noise intensity. As expected, for low
values of the correlation time we observe the same results than in
the white noise case. These effects are well visible in Fig.
\ref{aree} where the three curves show the nonmonotonic behavior
of the area of the maximum pattern as a function of noise
intensity. The interaction step here considered is 1400, which
correspond to the biggest pattern area found in our calculations.
The first curve ($\tau_c=1$) is quite the same found in the white
noise case. The value of maximum in the third curve ($\tau_c=100$)
is not so different from the previous one ($\tau_c=10$), because
its value is approaching the maximum possible value of 10.000 into
the used grid.

\begin{figure}[htbp]
 \begin{center}
  \includegraphics[height=9cm]{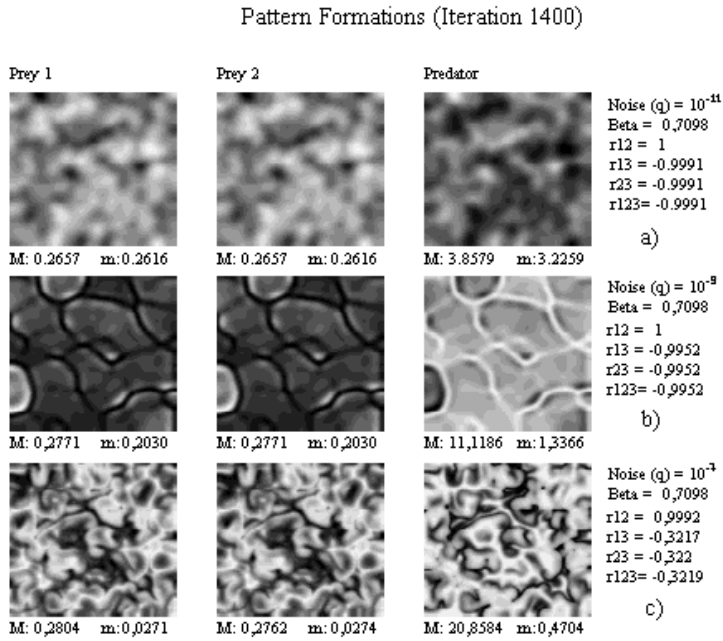}
  \vskip -0.3cm
\caption{Pattern formation for preys and predator with homogeneous
initial distribution, at time iteration $1400$ for $\tau_c=1$ and
noise intensity: $q = 10^{-11}, 10^{-9},10^{-7}$. $r_{12},
r_{13},r_{23},r_{123}$ represent respectively prey-prey,
prey1-predator, prey2-predator and total preys-predator
correlation. See the text for the values of the other parameters.}
 \vskip -0.8cm
  \label{pat}
 \end{center}
\end{figure}

The pattern formation is visible in Fig.~\ref{pat}, where we
report three patterns of the two preys and the predator for the
following values of noise intensity: $q=10^{-11}, 10^{-9},
10^{-7}$ and $\tau_c=1$.
 The initial spatial distribution
is homogeneous and equal for all species, that is
$x_{ij}^{init}=y_{ij}^{init}=z_{ij}^{init}= 0.25$ for all sites
($i,j$). We see that a spatial structure emerges with increasing
noise intensity. At very low noise intensity ($q = 10^{-11}$),
 the spatial distribution appears almost homogeneous without
strong pattern formation (see Fig.~\ref{pat}a). We considered here
only structured pattern, avoiding big clusterization of density
visible in the case of anticorrelated preys. At intermediate noise
intensity ($q = 10^{-9}$) spatial patterns appear. As we can see
the structure disappears by increasing the noise intensity (see
Fig.~\ref{pat}c). Consistently with Fig.~\ref{aree}, we find that
for higher correlation time $\tau_c$ the qualitative shape of the
patterns shown in Fig.~\ref{pat} are repeated, but with a shift of
the maximum area (darkest patterns) toward higher values of the
noise intensity.

\section{Conclusions}
The noise-induced pattern formation in a coupled map lattice of
three interacting species, described by generalized Lotka-Volterra
equations in the presence of multiplicative colored noise, has
been investigated. We find nonmonotonic behavior of the mean area
of the maximum patterns as a function of noise intensity for all
the correlation time investigated. For increasing values of the
correlation time $\tau_c$ we observe an increase of the area of
the pattern and a shift of the maximum value towards higher values
of the noise intensity. The nonmonotonic behavior is also found
for the area of the patterns as a function of the evolution time.
 \vskip 0.2cm
 This work was
supported by \mbox{INTAS Grant 01-0450}, by INFM and MIUR.

\end{document}